%% file: paper.tex
\documentclass[aps,pra,preprint,notitlepage,superscriptaddress]{revtex4-1}
\usepackage{graphicx}
\usepackage{amsmath}
\usepackage[latin1]{inputenc}
\usepackage{amsmath}
\usepackage{braket}
\usepackage{color}
\usepackage{bbm}
\usepackage{amssymb,amsfonts,latexsym,color,dcolumn,bm}
\usepackage{hyperref}

\usepackage{pgfplots}

\input{defs.tex}

\newcommand{\JSA}{\phi}
\newcommand{\fJSA}[1]{\funci{\JSA}{\nu,\eta}{#1}}
\newcommand{\dmJSA}{\mathbf{f}}

\newcommand{\fPM}[1]{\func{\Phi}{#1}}
\newcommand{\fJSAc}[1]{\func{\JSA^*_{\nu,\eta}}{#1}}
\newcommand{\MPQ}{\affiliation{Laboratoire Mat\'eriaux et Ph\'enom\'enes Quantiques, Universit\'e Paris Diderot, Sorbonne Paris Cit\'e, CNRS-UMR 7162, Case courrier 7021, 75205 Paris Cedex 13, France}}
\newcommand{\Klow}{K_\textrm{min}}
\newcommand{\Kexp}{K^\textrm{exp}_\textrm{min}}

\newcommand{\fFP}[2]{\funci{f}{#1}{#2}}
\newcommand{\FP}[1]{f_{#1}}

\newcommand{\figref}[2]{#1}

\begin{document}

\title{Direct high-resolution characterization of quantum correlations via classical measurements}

\author{A. Eckstein}
\MPQ

\author{G. Boucher}
\MPQ

\author{A. Lema{\^\i}tre}
\affiliation{Laboratoire de Photonique et Nanostructures, CNRS-UPR20, Route de Nozay, 91460 Marcoussis, France}

\author{P. Filloux}
\MPQ

\author{I. Favero}
\MPQ

\author{G. Leo}
\MPQ

\author{J. E. Sipe}
\affiliation{Department of Physics and Institute for Optical Sciences, University of Toronto, 60 St. George Street, Ontario M5S 1A7, Canada}

\author{M. Liscidini}
\affiliation{Department of Physics, University of Pavia, Via Bassi 6, I-27100 Pavia, Italy}

\author{S. Ducci}
\email{sara.ducci@univ-paris-diderot.fr}
\MPQ

\begin{abstract}Quantum optics plays a central role in the study of fundamental concepts in quantum mechanics, and in the development of new technological applications. Typical experiments employ non-classical light, such as entangled photons, generated by parametric processes. The standard characterization of the sources by quantum tomography, which relies on detecting the pairs themselves and thus requires single photon detectors, limits both measurement speed and accuracy. Here we show that the spectral characterization of the quantum correlations generated by two-photon sources can be directly performed classically with an unprecedented spectral resolution. This streamlined technique has the potential to speed up design and testing of massively parallel integrated sources by providing a fast and reliable quality control procedure. Adapting our method to explore other degrees of freedom would allow the complete characterization of biphoton states generated by parametric processes.\end{abstract}

\maketitle

\section{Introduction}

\begin{figure}[t]
\begin{center}
\includegraphics[width=\columnwidth]{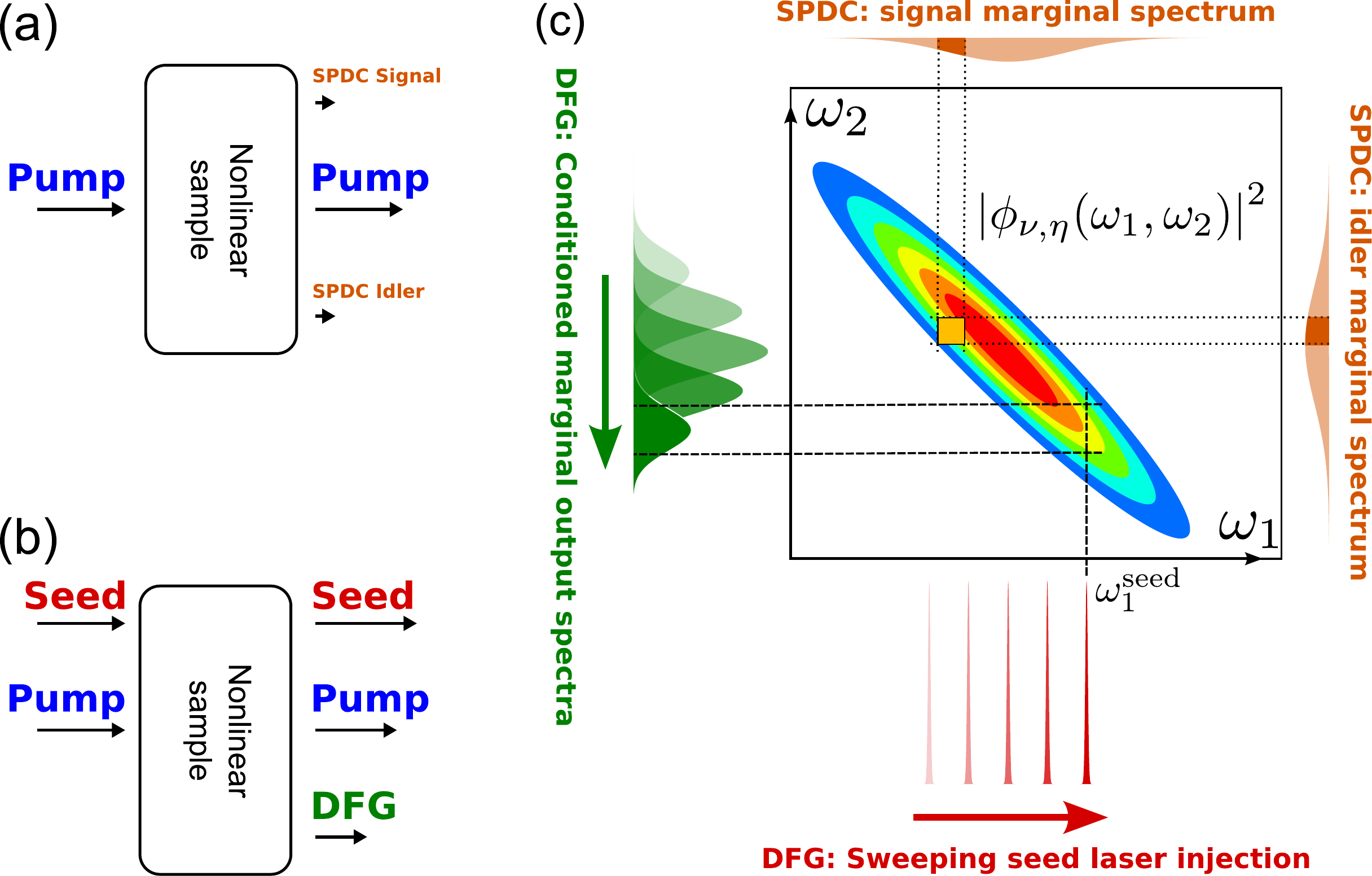}
\end{center}
\caption{Working principle:
(a) In a sample with a second order optical nonlinearity, SPDC probabilistically converts a photon from the coherent pump beam into a signal and idler photon pair.
(b) If a coherent ``seed'' beam is introduced into this nonlinear sample in either the signal or the idler photon's mode, a DFG process takes place and the conversion rate of pump photons is stimulated and increased by a factor proportional to the seed beam power.
(c) Reconstruction of a typical JSD, shared by the SPDC and the DFG process for the same nonlinear sample and the same pump configuration. In the SPDC case, the emitted signal and idler photons are analyzed with spectrometers. By single photon coincidence detection of each spectrometer's transmission, the intensity of a ``pixel'', corresponding to the joint transmission of both spectrometers' filter characteristics, is measured. The whole JSD can be reconstructed after collecting a sufficient number of events.
In the DFG case, a narrow-band seed laser beam at frequency $\omega_1^\textrm{seed}$ stimulates the emission of a spectrally pre-conditioned coherent output beam in the idler mode. This spectrum is proportional to a "slice"  of the JSD  corresponding to the injected wavelength (see Eq. \ref{eq:proportionality}). Sweeping the seed wavelength allows the reconstruction of the JSD.}
\label{fig:schematic}
\end{figure}

One parametric process used to generate quantum correlated photons is spontaneous parametric down-conversion (SPDC), the probabilistic conversion of a ``pump'' photon into a pair of photons, ``signal'' and ``idler'', in a nonlinear medium (see Fig. \ref{fig:schematic}a). In general, the probability of such an event is very small; thus the quantum state describing the radiation field in the frequency regime of signal and idler is mostly the vacuum state $\vac$, but also contains a normalized two-photon component with a small probability amplitude $\gamma$:
\begin{equation}
\ket{\psi_\textrm{pair}}=\frac{1}{\sqrt{2}}\sum_{\nu,\eta}
\iintd{\omega_1}{\omega_2}\,\funci{\JSA}{\nu,\eta}{\omega_1,\omega_2}\funcopdi{a}{\nu}{\omega_1}\funcopdi{a}{\eta}{\omega_2}\vac.
\end{equation}%
Here, $\abs{\gamma}^2$ is the probability with which a photon pair is emitted. $\nu$ and $\eta$ label the modes into which the photons are emitted, $\omega_1$ and $\omega_2$ indicate their frequencies. $\funci{\JSA}{\nu,\eta}{\omega_1,\omega_2}$ is the biphoton wavefunction, which characterizes all properties of the two-photon state and describes any quantum correlation between the two emitted photons. It is determined by the medium in which SPDC occurs, as well as the pumping scheme\cite{Yang2008}. In particular, $\abs{\funci{\JSA}{\nu,\eta}{\omega_1,\omega_2}}^2$ is known as the joint spectral density (JSD) and $\abs{\fJSA{\omega_1,\omega_2}}^2 \mathrm{d}\omega_1 \mathrm{d}\omega_2$ represents the probability of generating "photon 1" in the mode $\nu$ with frequency within $\mathrm{d}\omega_1$ of $\omega_1$, and "photon 2" in the mode $\eta$ with frequency within $\mathrm{d}\omega_2$ of $\omega_2$. The JSD is sufficient to estimate a lower bound for the degree of frequency entanglement between signal and 
idler (see appendix \ref{sec:error}).

So far, the JSD has been obtained by performing spectrally resolved single photon coincidence measurements\cite{Kim2005,Wasilewski2006,Avenhaus2009}. In practice this strategy is constrained by the pair generation probability, which must be much smaller than unity within the time resolution of the single photon detector, or an error would be introduced in the measured spectral correlations by the detection of simultaneously produced, but uncorrelated photons. Moreover, after a detection event the single photon detector has to be re-set to an operational state; this results in a deadtime $\tau_{D}$, which limits the maximally detectable coincidence rate to $\tau_{D}^{-1}$. These constraints lead to unavoidable limitations in the resolution with which the JSD can be determined. On the one hand, a large number of coincidences is required for reasonably low relative errors, demanding long integration times. On the other hand, short experimental runs are required to minimize any drift in the experimental 
conditions. Both of these requirements cannot be satisfied simultaneously, yet the emergence of advanced quantum optical applications in integrated devices\cite{OBrien2009,MartinLopez2012,Munro2012,Broome2013,Crespi2013,Collins2013} interfering multiple parametric photon sources necessitates a tool for their rapid characterization.

An alternative approach can be envisioned by recalling that, while SPDC can be described only in the framework of quantum theory, it can be viewed as difference frequency generation (DFG) in the quantum limit. Indeed, in DFG the conversion of pump photons to signal and idler pairs is stimulated by a seed beam, so SPDC can be considered as a DFG process stimulated by vacuum power fluctuations \cite{Helt2012}. The existence of a corresponding classical process naturally prompts one to question if it is possible to gain information about the quantum process by investigating only its classical analog. In the past, DFG has been used to determine the phase-matching function of SPDC sources\cite{Mason2002,Caillet2009}. It has also been experimentally demonstrated that seeded four-wave mixing (FWM) can be used to directly determine the number of pairs that would be generated by spontaneous FWM in ring resonators\cite{Azzini2012}. Theoretical studies have shown that DFG can be similarly used to determine the number 
of pairs that would be generated by SPDC, both in ring resonators and in other structures such as waveguides\cite{Helt2012}. In another context, DFG has been exploited for the realization of quantum cloning \cite{DeMartini2002}.

In this letter we take this classical-quantum connection even further and demonstrate experimentally that spectral quantum correlations of photon pairs that would be generated by SPDC can be investigated through the corresponding DFG measurements, despite the DFG process being describable completely in the framework of classical electromagnetic theory\cite{Liscidini2013}. Besides the intriguing fundamental aspect of this result, we show that our approach makes it possible to achieve very high resolution and increase data acquisition rates well beyond the state-of-the-art for spectrally resolved coincidence measurements\cite{Chen2009,Gerrits2011}. Finally, as one moves from SPDC to DFG, the increase of the produced output beam intensity by several orders of magnitude allows the replacement of the single photon detectors with an optical spectrum analyzer (OSA), a widely available general purpose instrument.

The strategy we employ is based on the fact that in a DFG experiment where the pump beam acting on a structure is the same as would be present in the corresponding SPDC experiment, but a quasi-CW signal beam is also introduced as a seed, the biphoton wavefunction $\fJSA{\omega_1,\omega_2}$ that \emph{would be relevant} in the spontaneous experiment plays the role of the response function of the structure that characterizes the generation of the stimulated light \cite{Liscidini2013}. In particular, the average number of photons stimulated in the mode $\eta$ with energy between $\omega_2$ and $\omega_2+\delta\omega_2$ by a coherent seeding beam exiting the system in mode $\nu$ and having energy centered at $\omega_1$ with a width of $\delta\omega_1$ can be written as
\begin{equation}
\begin{split}
\braket{\funcopdi{a}{\eta}{\omega_2}\funcopi{a}{\eta}{\omega_2}}_{\func{B_\nu}{\omega_1}}\delta\omega_2&\approx2\abs{ \func{B_\nu}{\omega_1}}^2\abs{\gamma}^2\abs{\fJSA{\omega_1,\omega_2}}^2\delta\omega_2\delta\omega_1\\
&=\abs{\func{B_\nu}{\omega_1}}^2\braket{\funcopdi{a}{\eta}{\omega_2}\funcopi{a}{\eta}{\omega_2}\funcopdi{a}{\nu}{\omega_1}\funcopi{a}{\nu}{\omega_1}}\delta \omega_2\delta\omega_1
\end{split}\label{eq:proportionality}
\end{equation}
where $\abs{\func{B_\nu}{\omega_1}}^2$ is the average number of photons in the coherent seeding beam, and where $\abs{\gamma}^2$ and $\braket{\funcopdi{a}{\eta}{\omega_2}\funcopi{a}{\eta}{\omega_2}\funcopdi{a}{\nu}{\omega_1}\funcopi{a}{\nu}{\omega_1}}\delta\omega_2\delta\omega_1$ are respectively the probability that a pair is generated, and the average number of pairs generated within $\delta\omega_2$ and $\delta\omega_1,$ by SPDC. Thus, by scanning the coherent seeding beam over the full spectrum (see Fig. \figref{1}{fig:schematic}), it is possible to obtain the JSD, $\abs{\fJSA{\omega_1,\omega_2}}^2$, that one would derive by coincidence measurements in a SPDC experiment. Moreover, Eq. \ref{eq:proportionality} tells us that the signal measured in the DFG experiments will be essentially $\abs{\func{B_\nu}{\omega_1}}^2$ times the one measured in the corresponding coincidence measurement, and thus several orders of magnitude larger.

\section{Experimental results}

\begin{figure}[t]
\begin{center}
\includegraphics[width=\columnwidth]{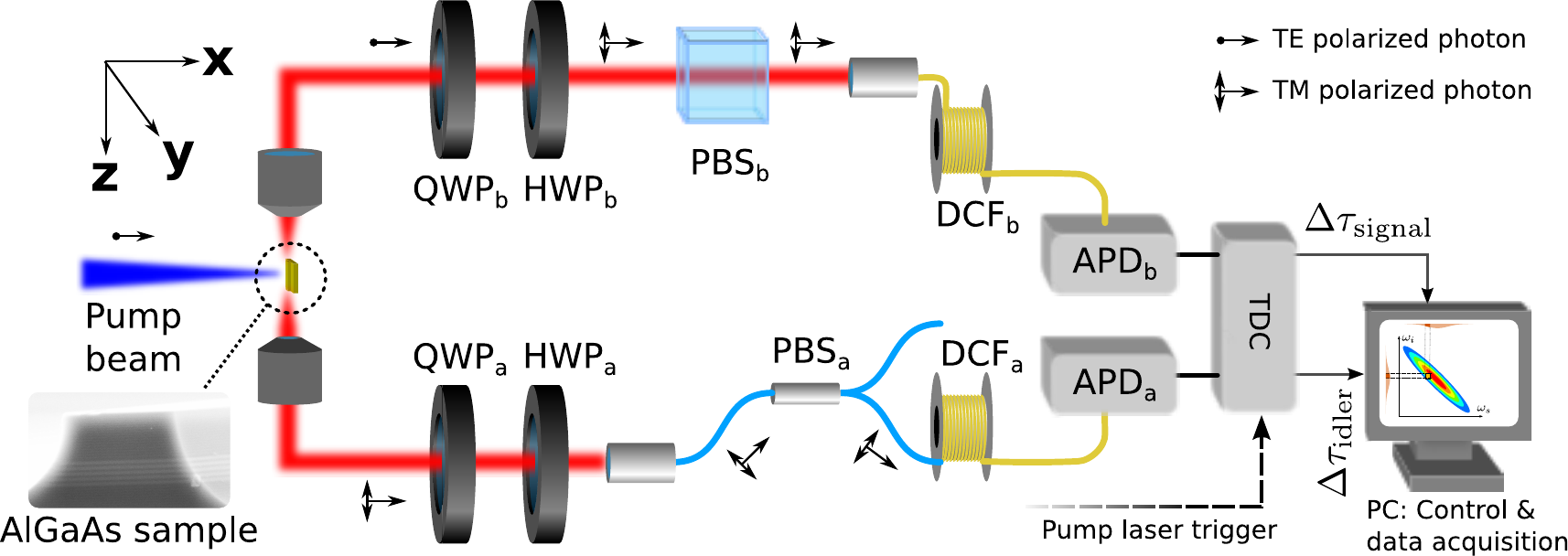}
\end{center}
\caption{Sketch of the experimental setup for SPDC-based spectral correlation measurements. A pump beam illuminates the top of the semiconductor device and couples into the waveguide so that counterpropagating pairs of signal and idler photons are created. One of the two possible type II phase-matched processes is selected with polarization optics (HWP$_{a/b}$, PBS$_{a/b}$) and two fiber single photon spectrometers\cite{Avenhaus2009} are used to analyze signal and idler photon. By introducing high group velocity dispersion with the fiber spools DCF$_{a/b}$, the arrival time of each photon at the avalanche photo-diodes APD$_{a/b}$ relative to the pump laser's electrical trigger signal reveals the photon frequency and is recorded by a personal computer via a time-to-digital converter (TDC).}
\label{fig:spdc_setup}
\end{figure}

To demonstrate the advantages of our characterization technique, we carried out SPDC and DFG-based frequency correlation characterizations of an integrated quantum light source: a picosecond-pulse-pumped AlGaAs ridge waveguide in a transverse pump configuration. An integrated microcavity with a resonance at the frequency of the pump beam enhances \cite{Caillet2009} the emission of counterpropagating photon pairs from two simultaneously phase-matched type II SPDC processes \cite{Orieux2013}.

In the SPDC experiment (Fig. \figref{2}{fig:spdc_setup}) we use a fiber spectrometer \cite{Avenhaus2009} to reconstruct the JSD by collecting one photon pair at a time. When a signal and an idler photons originating from the same trigger pulse are detected by the single photon detectors, the corresponding pair of signal/idler arrival times is added to a joint histogram. With a sufficiently large number of collected events this procedure yields the JSD, which is proportional to the coincidence counts plotted in Fig. \figref{4}{fig:results}a. Here, the spectral resolution is $\Delta\lambda_\textrm{SPDC}=\val{224}{pm}$, limited by the temporal jitter of the single photon detection signal relative to the pump trigger signal.

\begin{figure}[t]
\begin{center}
\includegraphics[width=\columnwidth]{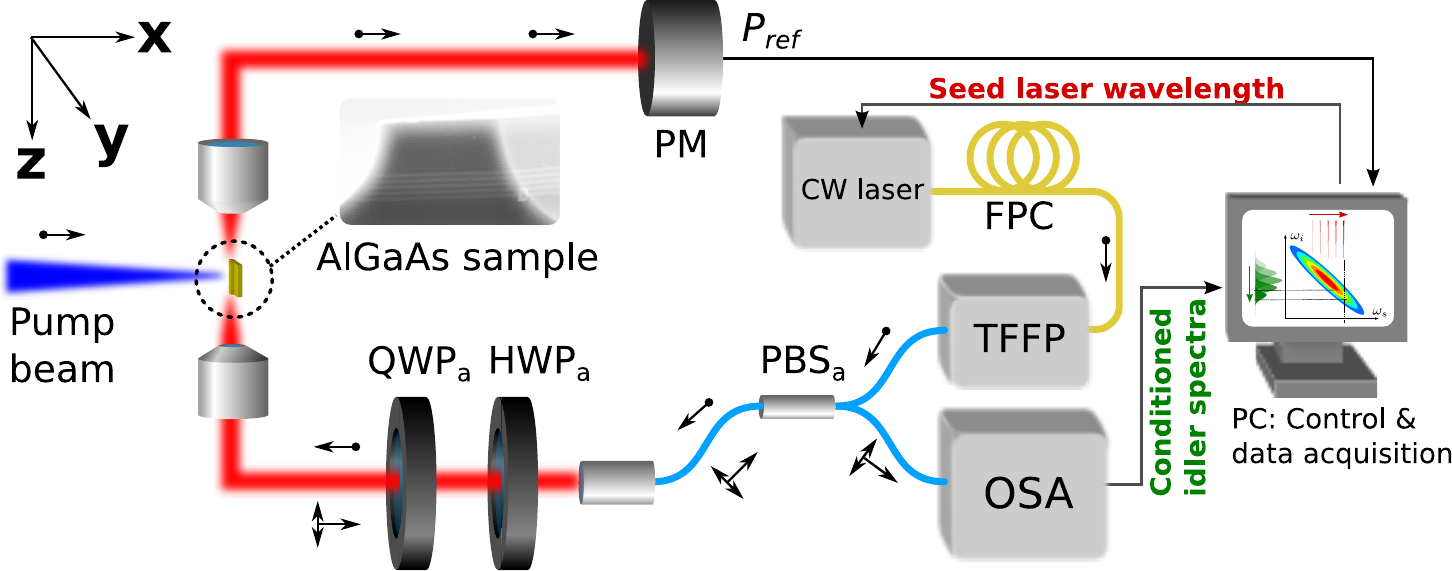}
\end{center}
\caption{Sketch of the experimental setup for DFG-based spectral correlation measurements. Beside the pump beam as in the SPDC experiment shown in Fig. \figref{2}{fig:spdc_setup}, we inject an additional CW seed laser into the waveguide's signal mode. Its polarization, adjusted by fiber polarization controller FPC and filtered by PBS$_a$, is used to select the same type II process as in the SPDC experiment.  The transmitted seed laser power $P_\text{ref}$ is measured by the powermeter PM. The tunable, fibered Fabry-Perot filter TFFP is used for spectral clean-up of the seed laser line. The backward emitted DFG beam has the same beam path as the seed beam, but has opposite polarization and propagation direction. The fiber integrated PBS$_a$ is therefore used as a combiner/splitter for both beams to retrieve the DFG output and guide it to an optical spectrum analyzer (OSA).}
\label{fig:dfg_setup}
\end{figure}

In the DFG experiment (see Fig. \figref{3}{fig:dfg_setup}) we collect the idler spectrum generated, under the same pumping condition, by sweeping a CW seed beam over the signal bandwidth of the spontaneous process. In accordance with Eq. \ref{eq:proportionality}, each pre-conditioned idler spectrum is proportional to the ``slice'' of the JSD corresponding to the pairs generated with the signal photon at the wavelength of the CW seed. The measurement result is presented in Fig. \figref{4}{fig:results}b. Its spectral resolution in signal axis is determined by the accuracy of the seed laser wavelength (\val{20}{pm}), while for the idler axis it is given by the OSA resolution (\val{20}{pm}).

\begin{figure}[t]
\begin{center}
\includegraphics[width=\columnwidth]{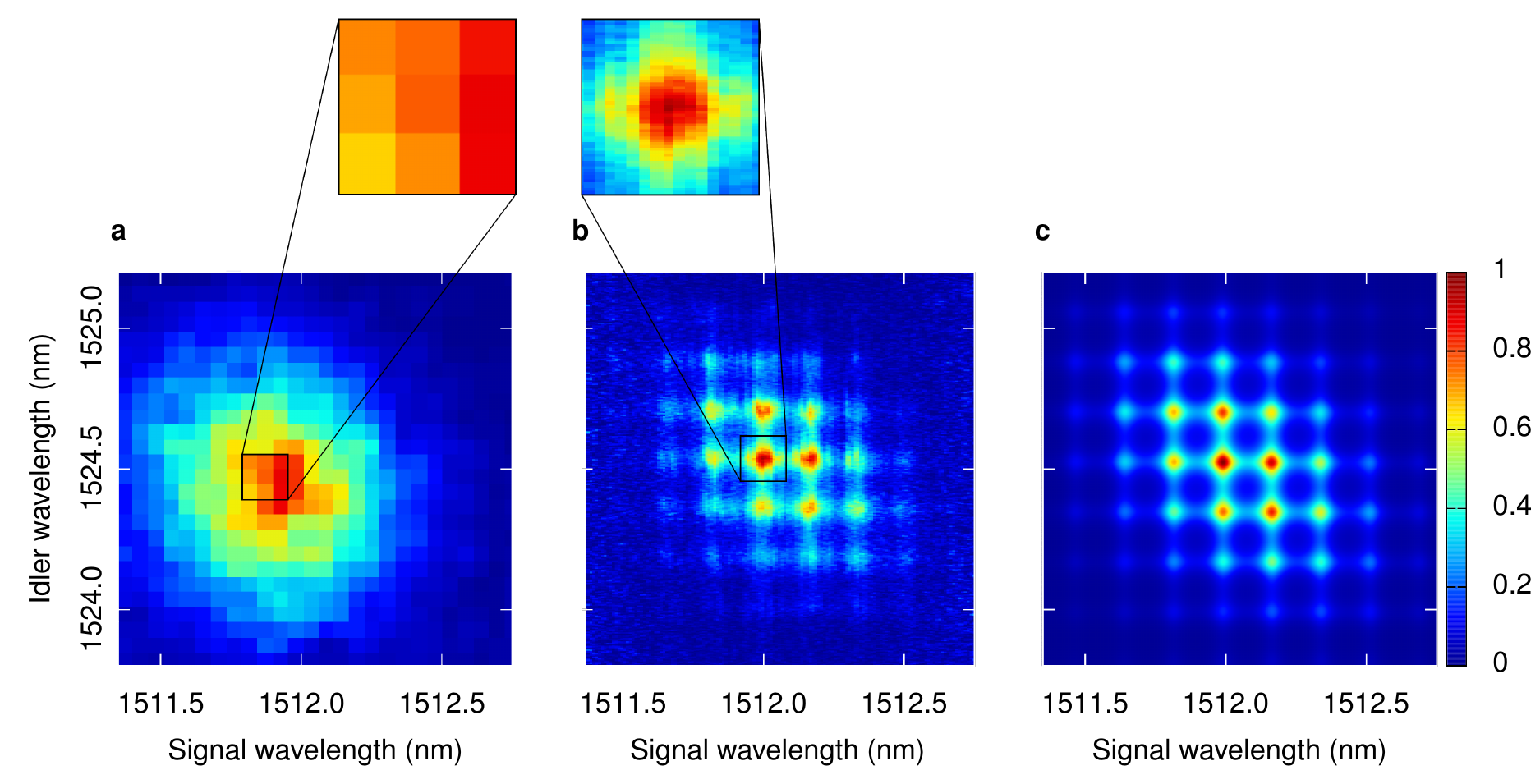}
\end{center}
\caption{Results: (a) Experimental JSD obtained by SPDC-based measurements (see Fig. \figref{2}{fig:spdc_setup}) with a sampling rate of $25 \times 25$ pixels over $\val{1.4}{nm}\times\val{1.4}{nm}$, an integration time of \val{120}{min}, and a spectral resolution of \val{224}{pm}. The pixel pitch of $\val{56}{pm}\times\val{56}{pm}$ is determined by the group velocity dispersion of the DCF coils and the temporal resolution of the TDC to measure photon arrival times.
(b) Experimental JSD obtained by DFG-based measurements (see Fig. \figref{3}{fig:dfg_setup}) with a sampling rate of $141 \times 501$ (non-quadratic) pixels over $\val{1.4}{nm}\times\val{1.4}{nm}$, an integration time of \val{45}{min}, and a spectral resolution of \val{20}{pm}. The pixel pitch of $\val{10}{pm}\times\val{2.8}{pm}$ corresponds to the scanning steps of the seed laser on the x-axis and to the spectral span over the number of data points of the OSA on the y-axis. The raw spectral data has been normalized to account for varying seed power (see appendix \ref{sec:methods}), leading to increased noise levels towards the left and the right edge of the plot. The visible offset between SPDC and DFG central wavelengths is caused by a shift of the central pump wavelength by \val{0.1}{nm} when re-locking the pump laser.
(c) Numerical calculation of the SPDC photon pairs' JSD generated by the device under study.}
\label{fig:results}
\end{figure}

In order to compare these results with the expected JSD, we theoretically calculated it with our experimental parameters (see Fig. \figref{4}{fig:results}c and appendix \ref{sec:sim}). The characteristic grid pattern is the result of the Fabry-Perot interferences due to the high index mismatch at the waveguide facets, which modify the vacuum power fluctuations within the waveguide. Interestingly, this effect was theoretically predicted for resonant SPDC devices \cite{URen2010} and has \textit{never} been observed, due to the limitations on the resolution of single photon detection-based measurements. The JSD obtained in the DFG measurement is to our knowledge the first to demonstrate it: theory and experiment are in excellent agreement. The high resolution JSD measurement boosts the pixel count over the SPDC results by two orders of magnitude while taking less than half as long to collect. Thanks to this dramatic increase in data acquisition rate, it becomes possible to fully exploit the spectral 
resolution of the seed laser and the detector and at the same time minimize statistical errors within realistic measurement durations. The Schmidt number $K=1.05$ (obtained from simulations) quantifies the spectral entanglement of SPDC photon pairs emitted by the sample\cite{Grobe1994,Law2000}. From the measured high resolution JSD, we can estimate an experimental lower boundary at $\Kexp=1.04$, with a theoretical value of $\Klow=1.03$ (see \ref{sec:error}).

\section{Conclusion}

In conclusion, we have implemented a novel technique to reconstruct the joint spectral density of biphoton states emitted by parametric processes based on a completely classical DFG experiment. It significantly out-performs spectrally resolved single photon coincidence measurements, both in terms of measurement time and resolution. The \val{20}{pm} spectral resolution achieved here can be improved by an order of magnitude using state-of-the-art spectrum analyzers and lasers. Already it is a qualitative advance over previous methods and reveals details of the biphoton JSD that could never have been observed before. We have demonstrated that this technique constitutes a fast, accurate and reliable tool for the characterization of photon pair sources. It opens the way to a new generation of experiments to explore hitherto unstudied aspects of nonclassical states of light.

\section*{Acknowledgments}

M.L. was supported by the Italian Ministry of University and Research, FIRB Contract No. RBFR08XMVY, by the Fundation Alma Mater Ticinensis. J.S. was supported by the Natural Sciences and Engineering Research Council of Canada. This work was partly supported by the French Brazilian ANR HIDE project and by R\'egion Ile de-France in the framework of C'Nano IdF with the TWILIGHT project. S.D acknowledges the Institut Universitaire de France.

\appendix

\section{Experimental methods}\label{sec:methods}

Figures \figref{2}{fig:spdc_setup} and \figref{3}{fig:dfg_setup} depict the measurement set-ups for the direct reconstruction of the JSD for SPDC and DFG-based methods, respectively. In each experiment, the sample is pumped by a mode-locked Coherent Mira Ti:Sapphire picosecond pulse laser at \val{759.1}{nm}, with a \val{0.4}{nm} spectral FWHM. Its repetition rate is reduced from \val{76}{MHz} to \val{3.8}{MHz} with an APE Pulse Select acousto-optical pulse picker introducing a temporal jitter of up to $\tau_\textrm{PP}=\val{200}{ps}$ into the pump laser's trigger signal.

The sample is a chemically-etched ridge AlGaAs waveguide grown by molecular-beam epitaxy. The employed phasematching scheme is non-collinear with a pump beam impinging on top of the waveguide at almost perpendicular incidence\cite{Caillet2009}. Signal and idler beam are emitted in cross-polarized, counterpropagating modes from either of two simultaneously phase-matched type II SPDC processes\cite{Orieux2013} (see appendix \ref{sec:sim} for details).

In SPDC measurements, both signal and idler photons are collected on either sides of the source with X40 microscope objectives. A set of wave plates and a polarizer allow us to select the correct polarization mode for each photon. Both photons then travel through DCF spools with a dispersion of $D_\textrm{DCF}=\val{-1475}{\frac{ps}{nm}}$. Free-running idQuantique id220 avalanche photo diodes (APD) act as single photon detectors at the end of each fiber. Detection efficiency is set to $20\%$, and the timing jitter of the electrical detection signal is $\tau_\textrm{APD}=\val{250}{ps}$. A quTools quTau TDC, connected to the APDs, measures the photons' arrival times relative to the pump laser's electrical trigger pulse with a $\tau_\textrm{TDC}=\val{81}{ps}$ mean temporal bin size. The spectral resolution of the fiber spectrometer assembly is given by the joint temporal jitter of the pulse picker and the APD, as well as the TDC time resolution over DCF dispersion, resulting in $\Delta\lambda_\textrm{SPDC}=\frac{
\sqrt{\tau_\textrm{PP}^2+\tau_\textrm{APD}^2+\tau_{TDC}^2}}{D_\textrm{DCF}}=\val{224}{pm}$.

In DFG measurements, we use a Yokogawa 6730C OSA with a resolution of \val{20}{pm} for spectral analysis of the idler beam. We employ a Tunics-Plus CW laser with a line-width of \val{100}{kHz} at an output power level of \val{8}{mW} to stimulate downconversion. With the help of the OSA, we detected a deviation from the nominal output wavelength by \val{-0.33}{nm} and verified its relative wavelength accuracy to be within the OSA resolution. The tunable fibered Fabry-Perot filter TFFP from ozOptics was used to clean the seed laser line; it has a Gaussian transmission profile with a \val{1.1}{nm} FWHM set to be centered at \val{1512.1}{nm}. The filter causes a variable seed laser power during the seed lasers wavelength sweep, which we monitored by recording the power $P_\textrm{ref}$ of the seed beam exiting the waveguide and accounted for in Fig. \figref{4}{fig:results}b by dividing the experimental value for each data point by the corresponding seed beam power.

\section{The nonlinear light source and the numerical calculation of the biphoton joint spectral density}\label{sec:sim}

\begin{figure}
\begin{center}
\includegraphics{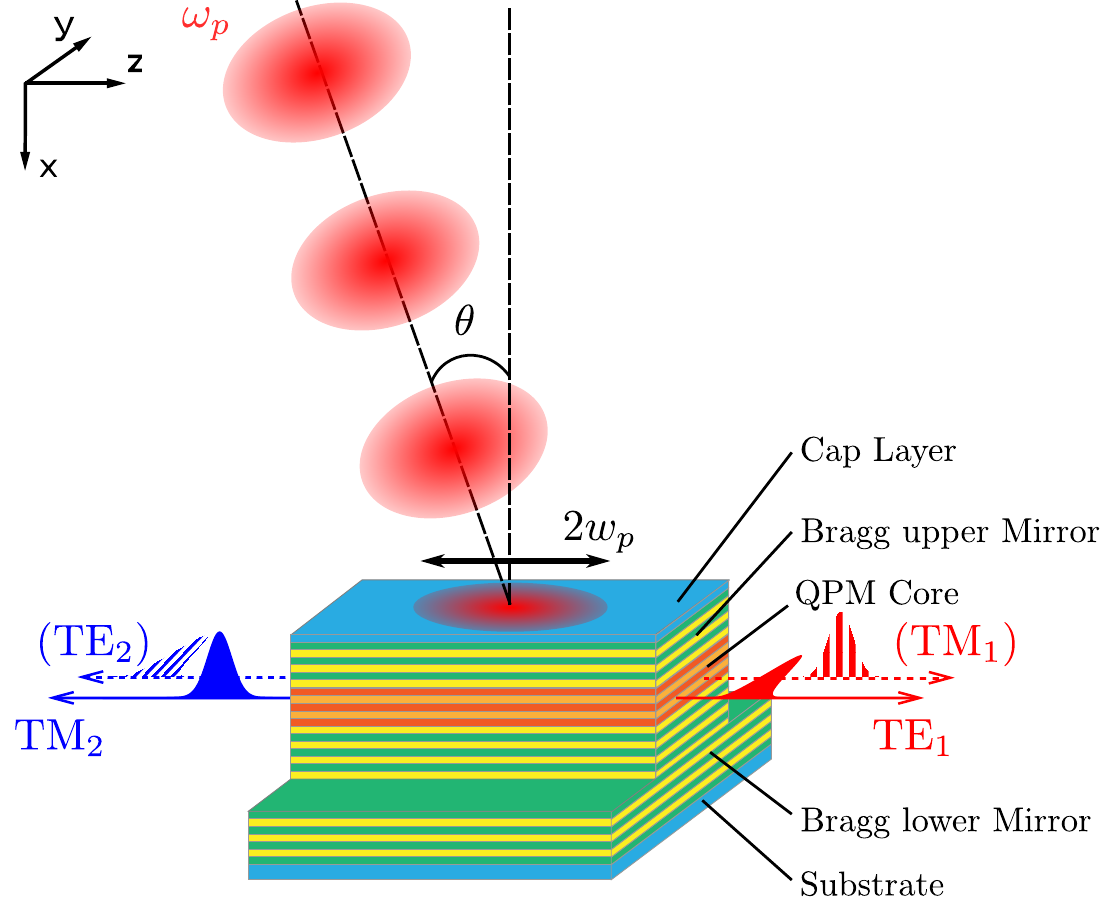}
\caption{Working principle of the two-photon source employed to test our technique: Pump laser pulses with a beam waist $w_p$ along the ridge direction z impinge on the top of the waveguide at an angle $\theta$. Photon pair TE$_1$ and TM$_2$ or photon pair TE$_2$ and TM$_1$ are emitted in opposite directions with orthogonal polarizations through two distinct type II SPDC processes.}
\label{fig:source}
\end{center}
\end{figure}

\begin{figure}
\begin{center}
\input{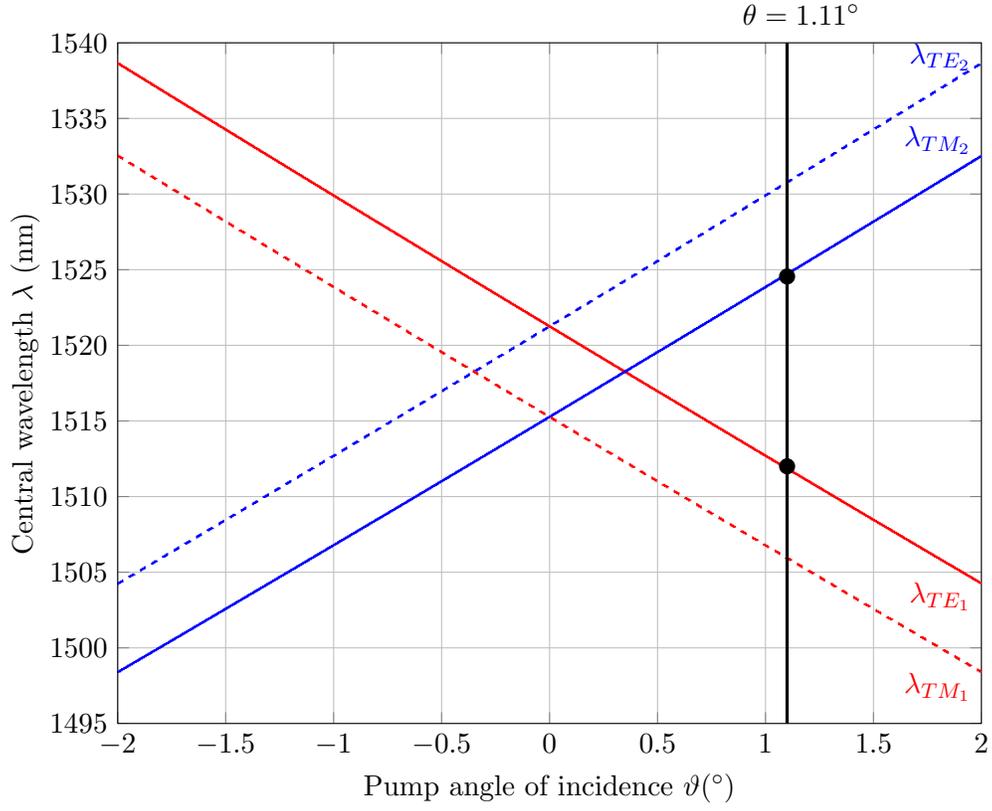}

\caption{Tunability curves of the counterpropagating source. The central wavelengths of the emitted photons are given as a function of the pump angle of incidence $\vartheta$. The pump angle used in the experiments is $\theta=1.11^\circ$, and the black circles mark the wavelengths of the photon pair TE$_1$, TM$_2$.}\label{fig:tuning}
\end{center}
\end{figure}

Fig.\ref{fig:source} is a sketch of the transversally pumped semiconductor photon pair source used in this experiment \cite{Caillet2009,Orieux2013}, a chemically-etched ridge AlGaAs waveguide grown by molecular-beam epitaxy. Two Bragg mirrors create a microcavity for the pump beam to enhance photon production. A quasi-phase matching core is implemented to compensate for the pump beam momentum in the vertical direction. Two type II process occur simultaneously; the tunability curves\cite{Orieux2013} as a function of the pump beam angle $\theta$ are shown in Fig. \ref{fig:tuning}. Polarization optics elements in our experiment allow us to select the interaction that produces the pair TE$_1$, TM$_2$, which we label ``signal'' and ``idler'', respectively.

The theoretical expression of the biphoton joint spectral density (JSD) function $|\fJSA{\omega_1,\omega_2}|^2$ generated by this device is:
\begin{equation}
|\fJSA{\omega_1,\omega_2}|^2 = \frac{1}{\mathcal{N}}
\abs{\func{\alpha_p}{\omega_1 + \omega_2}}^2
\abs{\fPM{\omega_1,\omega_2}}^2
\fFP{\textrm{M}}{\omega_1+\omega_2}\fFP{\textrm{TE}}{\omega_1}\fFP{\textrm{TM}}{\omega_2}.
\end{equation}

Here $\func{\alpha_p}{\omega_1 + \omega_2}$ is the spectral amplitude
of the pump beam, $\fPM{\omega_1,\omega_2}$ is the three-wave-mixing phasematching function, $\FP{\textrm{M}}$ is a function describing the Fabry-Perot effect of the microcavity on the pump beam, and $\FP{\textrm{TE}}$ and $\FP{\textrm{TM}}$ describe the effect of the reflexion on the waveguide facets for the generated TE and TM polarized photons, respectively. $\mathcal{N}$ is a normalization constant chosen so that $\iintd{\omega_1}{\omega_2}\ |\fJSA{\omega_1,\omega_2}|^2=1$.

In the following we detail the expression of each of these terms.
The pump beam is generated by a mode-locked Ti:Sapphire laser emitting pulses of hyperbolic secant spectral shape, centered at a wavelength of $\lambda_p=\val{759.1}{nm}$ with a bandwidth $\Delta\omega = 2\pi \times \val{84}{GHz}$, or $\Delta\lambda=\val{0.4}{nm}$ FWHM.
\begin{equation}
\func{\alpha_p}{\omega} = \Sech{\frac{\left(\omega-\frac{2 \pi c}{\lambda_p}\right)}{\Delta\omega} }
\end{equation}
The phasematching function $\fPM{\omega_1,\omega_2}$ is determined by the nonlinear waveguide's geometric and dispersion properties, and by the spatial intensity distribution of the pump beam (in our case a Gaussian beam):
\begin{equation}
\fPM{\omega_1,\omega_2} = \int_{-L/2}^{+L/2} \func{S}{z} e^{i\func{\dk}{\omega_1, \omega_2, \theta} z} dz
\end{equation}
with $L = \val{2.1}{mm}$ the waveguide length. The pump profile in the $z$-direction is $\func{S}{z}=\Exp{- \frac{z^2}{w_p^2}}$; the waist, measured with the knife-edge technique, is $w_p=\val{0.24}{mm}$ in the plane of the waveguide. The angle of incidence $\vartheta\equiv\theta=1.11^{\circ}$ of the pump beam on the waveguide was deduced from the measured signal and idler central wavelengths and the angular tunability curve of the device in Fig. \ref{fig:tuning}. The phase-mismatch function $\dk$ is given by
\begin{equation}
\begin{split}
\func{\dk}{\omega_1, \omega_2, \vartheta} &=  \func{k_p}{\omega_1+\omega_2}\sin \vartheta  - \func{k_{TE}}{\omega_1} + \func{k_{TM}}{\omega_2}\\
    &= \frac{1}{c}\left( \left(\omega_1+\omega_2\right) \sin \vartheta - \omega_1 \func{n_{TE}}{\omega_1} + \omega_2 \func{n_{TM}}{\omega_2} \right).
\end{split}
\end{equation}
The frequency-dependent effective refractive indices $n_{TE}$ and $n_{TM}$ are calculated with a standard transfer matrix method taking into account the nominal structure of the device\cite{Orieux2011} ($n_{TE} \approx 3.099$ and $n_{TM} \approx 3.086$). This leads to the waveguide end facets exhibiting a reflectivity of $R_\text{TE} = 26.7\%$, $R_\text{TM} = 24.7\%$.

As we have seen, the JSD expression is affected by the microcavity filtering the pump beam and by the waveguide partly reflective end facets constituting another cavity. This aspect, already treated in a collinear geometry \cite{URen2010}, has been adapted to our transverse pump configuration. The spectral transmission of the microcavity is described by the function
\begin{equation}
\fFP{\textrm{M}}{\omega}=\frac{1}{1+4\left(\frac{\omega-\omega_\textrm{M}}{\Delta\omega_\textrm{M}}\right)^2}
\end{equation}
with $\omega_\textrm{M}=\frac{2 \pi c}{\lambda_\textrm{M}}$ and $\Delta\omega_\textrm{M}=\frac{2 \pi c \Delta\lambda_{\textrm{M}}}{\lambda_\textrm{M}^2}$. For the sample used in this experiment, the resonance wavelength is $\lambda_\textrm{M} = \val{759.1}{nm}$ and the corresponding resonance width (FWHM) is $\Delta\lambda_\textrm{M} = \val{0.28}{nm}$ \cite{Orieux2011}.
The Fabry-Perot effect for the signal and idler beams is taken into account through the function $\FP{\mu}$ with $\mu \in \{\textrm{TE}, \textrm{TM}\}$:
\begin{equation}
\fFP{\mu}{\omega}=\frac{1}{1+F_\mu\Sin{\frac{L n_\mu}{c}(\omega - \omega_\mu)}^2}
\end{equation}
where
\begin{equation}
F_\mu=\frac{4 R_\mu}{1- R_\mu}
\end{equation}
is the cavity finesse and $R_\mu$ the reflectivities of the waveguide's facets calculated by 2D FDTD ($R_\textrm{TE}=26.7\%$ and $R_\textrm{TM}=24.7\%$), and $\omega_\mu\equiv\frac{2\pi c}{\lambda_\mu}$ is the frequency of a Fabry-Perot peak. The corresponding central wavelengths $\lambda_\textrm{TE} = \val{1511.99}{nm}$ and $\lambda_\textrm{TM} = \val{1524.53}{nm}$ used in Fig. 4c are determined from the DFG experimental data.

\section{The joint spectral density and extraction of a lower boundary for the Schmidt number from experimental data}\label{sec:error}

Any spectrally entangled biphoton state can be decomposed into a superposition of separable biphoton states with the help of the Schmidt decomposition\cite{Grobe1994,Law2000}
\begin{equation}
\fJSA{\omega_1,\omega_2}=\sum_n c_n \func{\psi_n}{\omega_1} \func{\varphi_n}{\omega_2}
\end{equation}
where $c_n$ are the real positive Schmidt parameters and $\{\left(\psi_n,\varphi_n\right)\}$ is a complete basis of orthonormal spectral function pairs, the Schmidt modes. The associated Schmidt number $K=\left(\sum_n c_n^4\right)^{-1}$ is a measure of frequency entanglement and can also be understood as an effective Schmidt mode number. A separable state features the minimal value $K=1$. The calculation of $K$ can also be expressed in terms of the normalized joint spectral amplitude (JSA) function $\JSA_{\nu,\eta}$\cite{Eberly2006}:
\begin{equation}
\begin{split}
\frac{1}{K}=
\iiiintd{\omega_1}{\omega_2}{\omega_1^\prime}{\omega_2^\prime}\ \fJSAc{\omega_1,\omega_2} \fJSAc{\omega_1^\prime,\omega_2^\prime} \fJSA{\omega_1,\omega_2^\prime} \fJSA{\omega_1^\prime,\omega_2}.
\end{split}\label{eq:schmidtnumber}
\end{equation}
Even though both measurements presented here sample the JSD function $\abs{\JSA_{\nu,\eta}}^2$ instead of the JSA, we can still estimate a lower boundary $\Klow$ for the Schmidt number $K$ from the experimental data\cite{Harder2013}, which can be easily shown. Applying the triangle inequality to Eq. \ref{eq:schmidtnumber} gives:
\begin{equation}
\begin{split}
\frac{1}{K}&=\abs{\frac{1}{K}}\\
&\leq\iiiintd{\omega_1}{\omega_2}{\omega_1^\prime}{\omega_2^\prime}\ \abs{\fJSA{\omega_1,\omega_2}} \abs{\fJSA{\omega_1^\prime,\omega_2^\prime}} \abs{\fJSA{\omega_1,\omega_2^\prime}} \abs{\fJSA{\omega_1^\prime,\omega_2}}\\
&=\frac{1}{\Klow}\approx \frac{\Tr{\left(\dmJSA^\dagger \dmJSA\right)^2}}{\Tr{\dmJSA^\dagger \dmJSA}^2}=\frac{1}{\Kexp}.
\end{split}\label{eq:kmin}
\end{equation}
where $\dmJSA$ is a real, non-negative $M\times N$ matrix representing the discretized, unnormalized modulus joint amplitude $C \abs{\fJSA{\omega_1,\omega_2}}$ within a certain spectral range, with $C$ an arbitrary positive constant. Since $\Klow$ is calculated from the modulus of the joint amplitude, it takes into account only frequency correlations, and all phase information of the state represented by $\JSA_{\nu,\eta}$ is lost. Any additional amount of phase correlation present will increase the actual Schmidt number $K$ relative to $\Klow$, thanks to the triangle inequality. The matrix indices of $\dmJSA_{m,n}$ are connected to the function's frequency arguments via
\begin{equation}
\begin{split}
\omega_1=\bar\omega_1+m \Delta\omega_1\\
\omega_2=\bar\omega_2+n \Delta\omega_2\\
\end{split}
\end{equation}
with $\bar\omega_1,\bar\omega_2$ the lower boundaries of the frequency window under scrutiny. $\Delta\omega_1$ and $\Delta\omega_2$ are the discretization constants which in experiment are identified with the scanning step width of the seed laser and the sampling pitch of the output spectrum measurement, respectively. It can be shown that the value for $\Klow$ does not depend on those constants, assuming that the variation of the joint spectral amplitude $\fJSA{\omega_1,\omega_2}$ within one discretization constant is negligible, and if the frequency range of the measurement covers most of the joint spectral amplitude:
\begin{equation}\int_{\bar\omega_1}^{\bar\omega_1+M\Delta\omega_1} \mathrm{d}\omega_1 \int_{\bar\omega_2}^{\bar\omega_2+N\Delta\omega_2} \mathrm{d}\omega_2 \abs{\fJSA{\omega_1,\omega_2}}^2\approx 1.\label{eq:totalintensity}\end{equation}
Just as the modulus of the joint amplitude can be obtained from the JSD by taking the square root, the matrix $\dmJSA$ can be calculated from the raw dataset of the DFG measurement:
\begin{equation}
\dmJSA_{m,n}=\sqrt{\frac{\mathbf{R}_{M-m,N-n}}{T_{M-m}}}\label{eq:measurementtojsa}
\end{equation}
where $\mathbf{R}_{m,n}$ is the measured intensity value for input wavelength bin $m$ and output wavelength bin $n$. The filter transmittance $T_m$ corrects for the optical loss at wavelength bin $m$ caused by the clean-up filter TFFP which is used to suppress broadband seed laser background. Note that in Eq. \ref{eq:measurementtojsa} we are implicitly assuming a linear dependency between wavelength and frequency, so that an equidistant set of data points in wavelength coordinates (which can be represented as a matrix because of this) is still equidistant in frequency domain. Although the relationship is in reality inverse rather than linear, the linearization is a very good approximation within the small spectral range of \val{1.4}{nm} at telecom wavelengths.

\begin{figure}
\begin{center}
\includegraphics{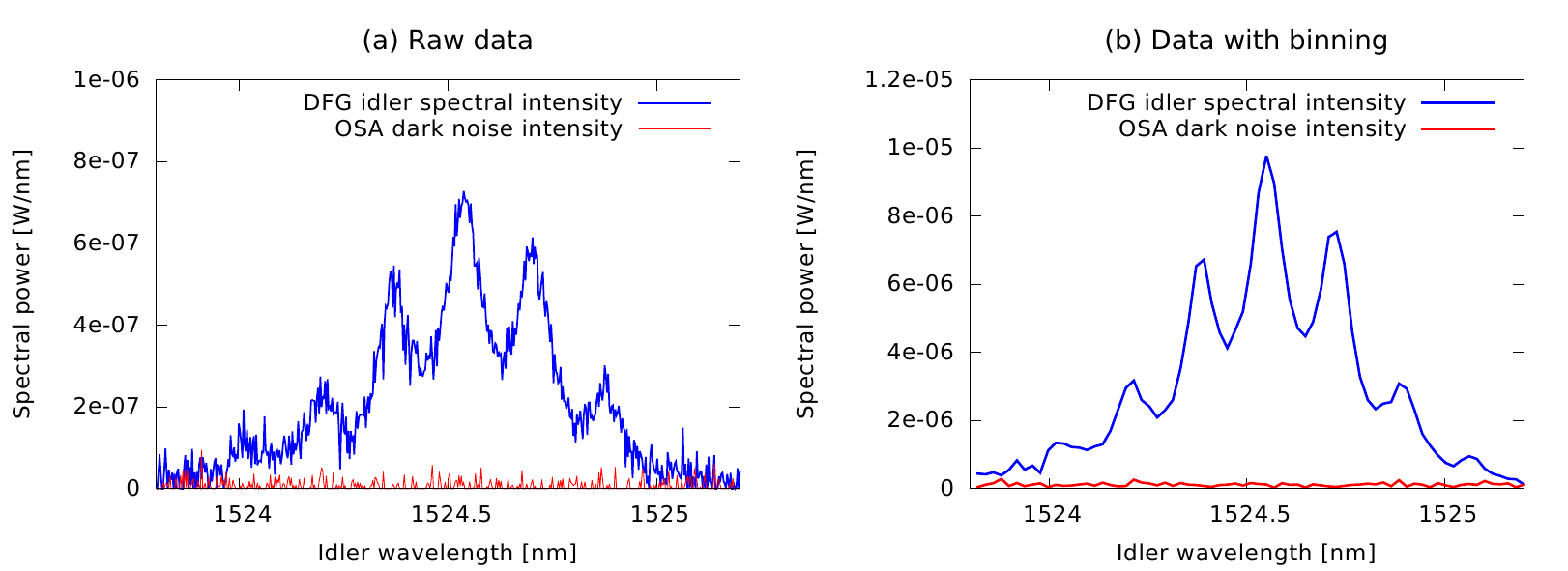}
\caption{Noise in the experimental data. (a) The DFG idler spectrum (blue) with a seed wavelength of $\lambda_\textrm{seed}=\val{1512.03}{nm}$ (corresponding to $\mathbf{R}_{63,n}$), and the dark noise (red) in a spectral measurement where the pump beam has been blocked out. (b) After applying a two-dimensional $2\times 7$ binning to the raw data matrix, the intensity noise is significantly reduced both in DFG idler spectrum (blue, corresponding to $\mathbf{R}^\prime_{32,n}$) and dark noise (red).}
\label{fig:noise}
\end{center}
\end{figure}

So far, we have not considered measurement noise, the main source of which is in our case the optical spectrum analyzer's intensity detection error, as is evident by the pixel-sized imperfections in the experimental data plot in Fig. 4b, and the obvious noise signature of the DFG intensity curve in Fig. \ref{fig:noise}a. Taking into account the full theoretical JSA with phase information for our experimental situation, we calculate a theoretical value of $K=1.05$, and for the theoretical lower boundary corresponding to Fig. 4c $\Klow=1.03$. We see that even in the presence of Fabry-Perot cavities for pump, signal and idler photon and its effect on the JSA's phase that $\Klow\approx K$. This is typically the case for picosecond-pulse pumped SPDC: The most important ways in which phase correlations can enter the JSA besides optical cavities are via a phase chirp of the pump pulses, which are typically minimized in mode locked lasers, and the group velocity dispersion of the nonlinear medium, which is small 
within the bandwidth of picosecond pulses\cite{URen2005}. We find however from raw experimental data $\Kexp=1.42$; this disparity is due to experimental noise and can be understood in terms of the Schmidt decomposition: when the perfectly smooth theory plot has a certain set of Schmidt coefficients $c_n$ that correspond to the occupation of the corresponding signal/idler mode pairs, the unavoidable discretization in a measurement situation and the addition of a random noise distribution that varies on a single pixel scale will artificially increase the occupation numbers of higher, faster oscillating Schmidt mode pairs $\left(\psi_n,\varphi_n\right)$, leading to higher values for the experimental lower boundary so that we expect not only $\Klow\approx\Kexp$ but also $\Klow\leq\Kexp$. Possible additional intensity noise sources on longer timescales are slow power fluctuations of the pump and the seed laser, or mechanical drift of the experimental setup due to ambient temperature changes or relaxation in opto-
mechanical components.

In Figure \ref{fig:noise}, we can see in both graphs that the signal-to-background ratio is lowest at minimal and maximal wavelengths. This problem is exacerbated by applying normalization against the filter function $T_m$. By reducing the data range taken into account for the calculation of the experimental value $\Kexp$, we decrease both actual information from the JSD and part of the background's contribution to the error of $\Kexp$, but by only excluding regions of the measurement data with a high relative background we improve the overall estimation. We verify this by comparing to the value $\Klow$ for a reduced theoretical dataset. Removing a ``frame'' with a width of \val{140}{pm}, we go from a data range of $\left[\val{1511.4}{nm};\val{1512.8}{nm}\right]\times\left[\val{1523.8}{nm};\val{1525.2}{nm}\right]$ to $\left[\val{1511.54}{nm};\val{1512.66}{nm}\right]\times\left[\val{1523.94}{nm};\val{1525.06}{nm}\right]$. The total intensity integral according to Eq. \ref{eq:totalintensity} decreases by $2\%$,
 and the theoretical lower boundary $\Klow$ decreases by $0.001$, so we consider the validity of the estimation to be preserved. The experimentally obtained value from the smaller data range is $\Kexp=1.22$, a significant improvement yet still far from the prediction.

To further suppress the noise of the experimental dataset, we introduce a binning and combine several single measurement values into one pixel:
\begin{equation}
\mathbf{R}^\prime_{m,n}=\sum^{b_x-1}_{k=0}\sum^{b_y-1}_{l=0}\mathbf{R}_{k+b_x m,l+b_y n}.
\end{equation}
With the binning constants $b_x=2$ and $b_y=7$ we obtain the dataset $\mathbf{R}^\prime$ with a pitch of $\val{20}{pm}\times\val{20}{pm}$ and $70\times71$ pixels, and Fig. \ref{fig:noise}b clearly shows an improvement of the signal-to-noise ratio of the DFG intensity.

By combining the data range reduction and the binning, we obtain $\Kexp=1.04$, which satisfies Eq. \ref{eq:kmin}. Applying the same treatment to the theoretical JSD dataset in Fig. 4c, we find a deviation of $\Klow$ by $-0.0011$. This indicates that we have introduced only a minor error due to reduction of the basis dataset, and that we have thus found a good estimate for the lower boundary of the Schmidt number $K$. Alternatively, a binning up to $(\val{30}{pm})^2$ pixel size and/or a choice of removing a spectral range of up to \val{180}{pm} from the edges of the dataset produces results consistent with $\Klow\leq\Kexp\leq K$. Beyond that, the information loss is too high and the resulting dataset's correlation drops below $\Klow$.

\input{paper.bbl}

\end{document}

%% file: defs.tex
\newcommand{\func}[2]{#1\!\left(#2\right)} 			
\newcommand{\funcrm}[2]{\mathrm{#1}\!\left(#2\right)} 		
\newcommand{\funci}[3]{{#1}_{#2}\left(#3\right)} 		

\newcommand{\unit}[1]{\mathrm{#1}} 				
\newcommand{\val}[2]{\ensuremath{#1\,\unit{#2}}} 		

\newcommand{\Exp}[1]{\textrm{e}^{#1}} 				

\newcommand{\Sech}[1]{\funcrm{sech}{#1}}

\newcommand{\Sin}[1]{\funcrm{sin}{#1}}

\newcommand{\abs}[1]{\left| #1 \right|}

\newcommand{\iintd}[2]{\iint\!\mathrm{d}#1\,\mathrm{d}#2 } 			
\newcommand{\iiiintd}[4]{\iiiint\!\mathrm{d}#1\,\mathrm{d}#2\,\mathrm{d}#3\,\mathrm{d}#4 } 	

\newcommand{\op}[1]{\ensuremath{\hat{\mathrm{#1}}}} 			

\newcommand{\funcopi}[3]{\op{#1}_{#2}\!\left(#3\right)} 		
\newcommand{\funcopdi}[3]{\op{#1}^\dagger_{#2}\!\left(#3\right)}	

\newcommand{\vac}{\ensuremath{\ket{0}}} 			
\newcommand{\Tr}[1]{\textrm{Tr}\!\left[#1\right]} 		

\newcommand{\dk}{\Delta k}

%% file: paper.bbl
%